\documentclass[doublecol]{epl2} 

\usepackage{microtype}
\usepackage[latin1]{inputenc}
\usepackage{mathtools}
\usepackage{amssymb}
\usepackage{mathtools}
\usepackage{bm}

\providecommand{\hm}{\pmb\bm}
\providecommand{\mbf}{\mathbf}

\providecommand{\what}{\hat}

\RequirePackage{xspace}
\newcommand*{\cf}{\emph{cf.}\xspace}

\newcommand*{\Eqn}[1]{equation~(\ref{#1})}
\newcommand*{\Eqns}[1]{equations~(\ref{#1})}

\newcommand*{\dalembert}{\square}

\newcommand*{\im}{{\mathrm{i}}}

\newcommand*{\mat}[1]{\mathsf{#1}}

\newcommand*{\w}{\omega}
\newcommand*{\Op}[1]{\what{\ten{#1}}}
\newcommand*{\op}[1]{\what{\mat{#1}}}
\newcommand*{\ten}[1]{\boldsymbol{\mathsf{#1}}}

\renewcommand*{\vec}[1]{\mbf{#1}}

\newcommand*{\unit}[1]{\hat{\vec{#1}}}

\newcommand*{\p}{\vec{p}}

\newcommand*{\grad}{\hm{\nabla}}
\newcommand*{\bdot}{\bm\cdot}
\renewcommand*{\div}{\grad\bdot}
\newcommand*{\cross}{\bm\times}
\newcommand*{\curl}{\grad\cross}

\title{Storming Majorana's Tower with OAM states of light in a plasma}
\shorttitle{OAM states in a plasma} 

\author{F.\,Tamburini\inst{1,2} 
\and B.\,Thid\'e\inst{3,4}
}
\shortauthor{F. Tamburini \& B.\,Thid\'e
}

\institute{                    
 \inst{1}%
 Department of Astronomy,
 University of Padua,
 vicolo dell' Osservatorio~3,
 IT-331\,22 Padua PD,
 Italy, EU
\\
 \inst{2}%
 Associazione CIVEN,
 via delle Industrie~5,
 TORRE HAMMON,
 IT-30175 Venezia-Marghera,
 Italy, EU
\\
 \inst{3}%
 Swedish Institute of Space Physics,
 {\AA}ngstr\"{o}m Laboratory,
 P.\,O.~Box~537,
 SE-751\,21 Uppsala,
 Sweden, EU
\\
 \inst{4}%
 Galilean School of Higher Education,
 University of Padua,
 Via Giovanni Prati~19,
 IT-351\,22 Padua PD,
 Italy, EU
}


\abstract{We extend the relationship between mass and spin angular
momentum, described by the bosonic spectrum of positive definite
mass particles of the Majorana solution to the Dirac equation, to
photons that acquire an effective Proca mass through the Anderson-Higgs
mechanism \cite{Anderson:PR:1963} when they propagate in a plasma. In an
earlier paper \cite{Tamburini&al:EPL:2010} we showed that if the plasma
is structured, it can impart orbital angular momentum (OAM) to the
photons that reduces the total Proca photon mass. Here we show, through
a generalisation of Majorana's solution, that photons with OAM in a
plasma cannot assume negative squared mass states. This means that there
exist interesting analogies with Quantum Gravity or General Relativity
models involving a modified action of the Lorentz group.}

\begin{document}

\maketitle

\section{Introduction}

In 1932, Majorana \cite{Majorana:NC:1932} formulated an alternative
solution to the Dirac equation for bosonic and fermionic
relativistic particles with null or positive definite rest mass
whose particle spectrum is also known as the ``Majorana Tower''
\cite{Magueijo:Book:2010}. This particular solution was formulated in
an attempt to avoid the problem of the negative squared mass solution
emerging from the relativistic electron wavefunction, which led to the
concepts of anti-electron and anti-matter, identified experimentally
by C.~D.~Anderson \cite{Anderson:PR:1933}. The particle spectrum 
described by the Majorana solution is characterized by a specific
mass/spin relationship that does not possess any correspondence with
the spectrum of known elementary particles in the Standard Model
\cite{Hagiwara&al:PRD:2002}. Instead, a subset of the Majorana
particle spectrum describes a particular class of fermions, the
so-called Majorana neutrinos. These fermions are supposed to be found
in neutrino-less double beta decays through a lepton-number violating
nuclear decay process \cite{Doi&al:PTPS:1985,Mohapatra&al:RPP:2007,
Avignone&al:RMP:2007}. In the spectrum of Majorana's solution, a
particle and its antiparticle are not mutually distinguishable
\cite{Majorana:NC:1937}.

Photons can in all aspects be considered to be Majorana particles.
They carry energy, linear momentum and angular momentum and have, in
free space, zero rest mass. Their total angular momentum $\vec{J}$ is
the sum of the spin angular momentum (SAM) $\vec{S}$ and the
orbital angular momentum (OAM) $\vec{L}$. From a classical
point of view, the electromagnetic field can carry linear
momentum as well as angular momentum all the way to infinity
\cite{Jackson:Book:1999,Thide:Book:2011} and therefore these physical
observables can be utilised for wireless information transfer over long
distances \cite{Gibson&al:OE:2004,Tamburini&al:APL:2011}.

A photon propagating in a plasma acquires an effective
mass $m^{\ast}$ due to its interaction with plasmons;
this effect is described by the Anderson-Higgs mechanism
\cite{Schwinger:PR:1962,Anderson:PR:1963} or as a hidden gauge
invariance of the Proca-Maxwell equations \cite{Mendonca:Book:2001}.
If the plasma is homogeneous and characterized by its unperturbed
plasma frequency $\w_{p0}$, and if the perturbation induced by the
photons in the plasma is neglected, the mass acquired by photons will be
$m^{\ast} =\hbar\w_{p0}/c^2$, where $\hbar$ and $c$ are the reduced
Planck constant and the speed of light, respectively.
In order to simplify the notation, we will henceforth in this Letter use
natural units, $e=c=\hbar=G=1$, so that the virtual photon mass in an
unperturbed, homogeneous plasma coincides with the unperturbed plasma
frequency.

Recently it was shown theoretically that a photon propagating in
a structured plasma will acquire OAM and an associated additional
effective mass \cite{Tamburini&al:EPL:2010}. This is because the OAM
acts as an effective mass-reducing term. An electromagnetic wave
traversing a plasma introduces a perturbation in the plasma particle
distribution, assumed to comprise electrons and heavy, immobile positive
ions providing a neutralising background. The square of the plasma
frequency becomes ${\w_p^2(x)=\w^2_{p0}\big(1+\eta(r,\varphi,z)\big)}$,
where $\eta(r,\varphi,z)$ is the electron distribution perturbation.
Then the following relationship between the acquired effective Proca
mass $\mu_\gamma$ and the perturbation in the motion of the electrons
caused by the electromagnetic (EM) wave associated with the photon
holds:
\begin{align}
\label{eq:mass}
 \mu_\gamma=\frac{1}{1+\frac{\unit{v}\cdot\grad\Phi}{E}}
 \Big(\w_p^2-4\pi\frac{n(r)\delta\dot{v}
 -\unit{v}\bdot\dalembert\grad\Phi}{E}\Big)^{1/2}
\end{align}
Here, $\unit{v}=\vec{v}/|v|$ denotes the unit direction
vector of the electron velocity in the plasma, $\vec{v}$,
assumed, in accordance with Ref.~\citenum{Tamburini&al:EPL:2010},
to be parallel to the electric field $\vec{E}$, and
${\delta\dot{v}=\unit{v}\bdot\partial_t{\vec{v}}=|\partial_t{v}|}$.

The EM wave propagating in the plasma is assumed to generate a
weak perturbation $\delta\vec{v}$ of the unperturbed electron
velocity vector $\vec{v}_0$ such that the perturbed velocity is
$\vec{v}=\vec{v}_0+\delta\vec{v}$, and higher order corrections can
be neglected. Let us consider the simple case of a helicoidal plasma
density perturbation $\tilde{n}\cos(\ell\varphi+q_0z)$ where $q_0$ is
the helix step. For the OAM value $\ell$ in the Proca squared mass term,
we then find that
\begin{multline}
\label{eq:mu_gamma^2}
\mu_\gamma^2
 = \frac{E}{E+\unit{v}\bdot\grad\Phi}\w^2_{p0}\big(1+\eta(r,\varphi,z)\big)
\\
 -\frac{1}{E+\unit{v}\bdot\grad\Phi}\Big(
 4\pi\delta \dot{ v}\big(n_0+\tilde{n}\cos(\ell\varphi+q_0z)\big)
\\
 -4\pi\vec{v}\bdot\dalembert\grad\Phi\Big)
\end{multline}
where $n_0$ is the unperturbed electron density and $m_e$ the electron
mass. The geometry of the perturbation is described in cylindrical
coordinates, $\vec{r}\equiv(r,\varphi,z)$ and the spiral formed plasma
perturbation winds along the $z$ axis, along which the EM wave is
propagating. Of course, one can describe the case of a more general
perturbation $f(\ell,q)$ as a superposition of helicoidal perturbations.
It is interesting to note that a similar behaviour is found in the
London equations that describe fields and currents in superconductors.
Also there, a characteristic length that defines the behaviour of EM
field is present \cite{Tinkham:Book:2004}. The London penetration
depth $\lambda_L$ is the characteristic length of penetration of a
magnetic field in a superconductor that depends on mass, density and
charge of the electrons in the super-conducting medium similarly to what
is seen in \Eqn{eq:mu_gamma^2}.

The term in \Eqn{eq:mu_gamma^2} containing the photon OAM $\ell_0$
acts as a \emph{negative squared mass term} for the virtual (Proca)
photon mass, exhibiting the possibility of generating a photon with
negative squared mass, as expected in certain Lorentz and gauge
symmetry violations \cite{Ruhl:NC:1970,Goldhaber&Nieto:RMP:1971,
Goldhaber&Nieto:RMP:2010,Ferrero&Altschul:PRD:2009}.

By extending the Majorana solution to the Dirac equation in free space
to photons in a plasma carrying OAM, we show in this Letter that the
Proca mass of a photon in a plasma is always positive definite when
Lorentz invariance is preserved.

\section{Space-time symmetries}

In the Proca-Maxwell equations, Lorentz invariance is preserved but the
gauge invariance is lost because both the vector and scalar potentials
of the EM field, $\vec{A}$ and $\Phi$, respectively,
become at all effects observable quantities. The EM energy density,
$u$, depends explicitly on the potentials and, consequently, on
the energy densities $\mu_\gamma^2\Phi^2/8\pi$ and $\mu_\gamma^2
\vec{A}^2/8\pi$ in the following way:
\begin{equation}
u = \frac{1}{8 \pi}\left(\mathrm{\textbf{E}}^2 + \mathrm{\textbf{B}}^2
 + \mu_\gamma^2\Phi^2
 + \mu_\gamma^2\mathrm{\textbf{A}}^2\right).
\end{equation}

Following the Majorana-Oppenheimer formulation of QED
\cite{Tamburini&Vicino:PRA:2008,Giannetto:LNC:1985},
we represent the photon wavefunction in Riemann-Silberstein
(RS) formalism, in which the wavefunction is described
by the complex RS vector ${\vec{G}=\vec{E}\pm\im\vec{B}}$
\cite{Silberstein:AP:1907a,Silberstein:AP:1907b,Berry:JOA:2004a,
Thide:Book:2011}. The photon obeys a Dirac-like equation with
the well-known problems of photon localisability
\cite{Mignani&al:LNC:1974,Bialynicki-Birula:APP:1994,
Bialynicki-Birula:PRL:1998,Thide:Book:2011}. Without loss of generality
for any particular choice of helicity state or multiplicative constants,
the Maxwell equations in free space become
\begin{subequations}
\begin{gather}
\label{eq:max1}
\im\curl{\vec{G}}=\pm \frac{\partial\vec{G}}{\partial t}
\\
\label{eq:max2}
 \div{\vec{G}}=0
\end{gather}
\end{subequations}
By using the correspondence principle
$\vec{p}\leftrightarrow\Op{p}\equiv-\im\grad$,
\Eqn{eq:max1} leads to a Dirac-like wave equation
\begin{align}
 \mp\im\frac{\partial}{\partial t}\vec{G}
 +\im{\Op{p}}\cross\vec{G}=\vec{0}
 \end{align}
whereas \Eqn{eq:max2} describes the transversality of the
EM fields with respect to the propagation direction, namely,
$\Op{p}\cdot\vec{G}=0$.

Let us consider the infinitesimal Lorentz transformations
described by the $4\times4$ matrix sets
$\ten{S}=\{\mat{S}_x,\mat{S}_y,\mat{S}_z\}$ and
$\ten{T}=\{\mat{T}_x,\mat{T}_y,\mat{T}_z\}$, where the set elements are
\begin{subequations}
\label{eq:lorentz1}
\begin{align}
\mat{S}_x
 = \begin{pmatrix}
    0&0&0&0\\
    0&0&0&0\\
    0&0&0&-1\\
    0&0&1&0
   \end{pmatrix}
\\
\mat{S}_y
 = \begin{pmatrix}
    0&0&0&0\\
    0&0&0&1\\
    0&0&0&0\\
    0&-1&0&0
   \end{pmatrix}
\\
\mat{S}_z
 = \begin{pmatrix}
    0&0&0&0\\ 
    0&0&-1&0\\
    0&1&0&0\\
    0&0&0&0
   \end{pmatrix}
\end{align}
\end{subequations}
and
\begin{subequations}
\label{eq:lorentz2}
\begin{align}
\mat{T}_x
 = \begin{pmatrix}
            0&1&0&0\\ 
            1&0&0&0\\
            0&0&0&0\\
            0&0&0&0
   \end{pmatrix}
\\
\mat{T}_y
 = \begin{pmatrix}
            0&0&1&0\\
            0&0&0&0\\
            1&0&0&0\\
            0&0&0&0
   \end{pmatrix}
\\        
\mat{T}_z
 = \begin{pmatrix}
            0&0&0&1\\ 
            0&0&0&0\\
            0&0&0&0\\
            1&0&0&0
   \end{pmatrix}
\end{align}
\end{subequations}
respectively. By taking the $3\times3$ spatial parts, one defines the
sub-matrices $\ten{s}=(\mat{s}_x,\mat{s}_y,\mat{s}_z)$ of the $\ten{S}$
matrices, and by writing the Hamiltonian of the Dirac-like equation for
the photon as (\cf Refs.~\citenum{Thide:Book:2011,Mignani&al:LNC:1974}
and \citenum{Bialynicki-Birula:PO: 1996,
Bialynicki-Birula:CQO7:1996,Kobe:FP:1999})
\begin{align}
 \op{H}=\pm\ten{s}\cdot\Op{p}
\end{align}
one obtains
\begin{align}
\label {Dirac}
 \im\frac{\partial}{\partial t}\vec{G} = \op{H}\vec{G}.
\end{align}
The Hamiltonian $\op{H}$ has eigenvalues $\pm{p},0\,$; the eigenvalue $0$ is
forbidden by the transversality condition \cite{Berestetskii&al:Book:1989}. 

\section{The Majorana Tower}

Let us now in more detail analyse the Majorana formulation of the
infinite-spin component Dirac equation with positive definite
real mass eigenvalues and in the absence of an external field
\cite{Majorana:NC:1932}, also known as the ``Majorana Tower'' of
particles \cite{Magueijo:Book:2010}. Here, because of the chosen
space-time symmetries of the Lorentz group applied to the Dirac equation
in free space, the mass spectra of bosons and fermions always assume
positive definite values and follow a precise mass/spin angular momentum
relationship.

A particular case is the class of zero rest mass particles such as
photons, neutrinos and gravitons that differ from their antiparticles
only because of their helicity state \cite{Tamburini&Vicino:PRA:2008};
they belong to a wider class of inequivalent relativistic
representations of massless vector fields with arbitrary helicity states
\cite{Fushchich&Nikitin:Book:1987}. 

Let us consider the particular case of the EM field. Conformal
invariance and the postulate of the vector character of the EM field
imply that part of these general fields do admit Maxwell's equations
as a special case. In these terms, each of these more general fields
can be described by a set of matrices $S_{\mu\nu}$ in a vectorial
representation of the generators $D(0,1)$ and $D(1,0)$, or by the
equivalent spinorial representations of the generators $D(1/2,0)$ and
$D(0,1/2)$ \cite{Tamburini&Vicino:PRA:2008}.

We write the general Dirac equation in the form
\begin{equation}
\label{eq:dirac}
\left[\gamma_0+W+\alpha\cdot\Op{p}-\beta\mu_\gamma\right]\vec{G}=0,
\end{equation}
where $\gamma_0$, $\alpha$ and $\beta$ are the usual Dirac matrices,
$\mu_\gamma$ is the mass of the particle and $W$ the energy
\cite{Dirac:PRSA:1928,Majorana:NC:1932,Thaller:Book:2010}. As is well
known, the Dirac equation for the relativistic electron admits solutions
with both positive and negative mass eigenvalues.

Let us apply the Majorana formalism to the photon in a perturbed plasma.
When a quantum of light is propagating in a region filled with plasma,
it acquires a positive virtual mass for both helicity states of the
wave equation. For this reason, the eigenvalues of the matrix $\beta$
of \Eqn{eq:dirac} and the quantity $\tilde{\vec{G}}\beta\vec{G}$ must
be positive definite. Through non-unitary transformations one can,
according to Ref.~\citenum{Majorana:NC:1932}, write the quantity
$\tilde{\vec{G}}\beta\vec{G}=\tilde{\vec{g}}\vec{g}$ of the RS
vector and obtain an equivalent formulation
\begin{align}
(\gamma_0 + W + \gamma\cdot\Op{p} - \mu_\gamma)\vec{g}=0
\end{align}
where the invariance of the function $\tilde{\vec{g}}\vec{g}$ is
provided by unitary transformations \cite{Majorana:NC:1932}.

From the infinitesimal Lorentz transformations, \Eqns{eq:lorentz1}
and (\ref{eq:lorentz2}), one writes a set of Hermitean operators,
$a_j=\im\mathsf{S}_j$ and ${b_j=-\im\mathsf{T}_j}$, for the $j$th
space-time coordinate. The commutation rules then become
\begin{alignat}{3}
\label{eq:commut}
[a_x, a_y] &= \im a_z, \quad &[b_x, b_y] &= -\im a_z 
\notag
\\
[a_x, b_x] &= 0, &[a_x, b_y] &= \im b_z, \quad [a_x, b_z] = \im b_y
\end{alignat}

The spectrum of these particles exhibits a relationship between the
intrinsic spin angular momentum, $s$, with another parameter, $m^\ast$,
the ``Majorana-mass'' term, which is related to the particle's rest
mass or to the virtual mass. This property is valid for both bosonic
($j=1,2,3,\ldots$) and fermionic ($j=1/2,3/2,5/2,\ldots\,$) solutions,
\begin{equation} 
\label{eq:majmass}
 \mu_\gamma=\frac {m^\ast}{s+1/2}
\end{equation}
and describes a spectrum cast in an infinite tower of particles with
positive mass values that decrease when the intrinsic angular momentum
of the particle increases.

\section{Space-time symmetries and plasma}

We now show that photons in a structured plasma follow a similar
mass/angular momentum relationship, with the difference that
the mass term is given by the Proca mass and the spin angular
momentum $\vec{S}$ is replaced by the total angular momentum,
$\vec{S}\rightarrow\vec{J}=\vec{L}+\vec{S}$.

During the propagation of a light beam, the total angular momentum
$\vec{J}$ is conserved \cite{Thide:Book:2011}. When photons traverse
a structured inhomogeneous medium, angular momentum conversions occur
with the result of generating beams of light carrying OAM. According
to Ref.~\citenum{Tamburini&al:EPL:2010}, one finds that the higher the
value of the OAM acquired by the photon, the lower the total effective
Proca mass of the particle, analogously to what is described with spin
angular momentum and mass for a Majorana particle. More precisely, in
the Proca mass formulation, the unperturbed squared mass quantity,
$m^{\ast2}$, becomes that of the perturbed plasma, $\mu_\gamma^2$, given
by \Eqn{eq:mu_gamma^2}.

In the Proca mass term the OAM term $\ell$ induced by the plasma
structure becomes evident with the characteristic scale $q_0$,
\begin{align}
\mu_\gamma =  \w_{p0}A\sqrt{1-
 \frac{4\pi\delta\dot{v}}{E+\unit{v}\bdot\grad\Phi}\,
  \frac{n_0+\tilde{n}f(\ell,q)}{ \w_{p0}^2 A^2}
 }
\end{align}
After some algebra, one finds that the Proca mass equations leads to
an approximate mass/angular momentum relationship that extends the
vacuum (free space) solution proposed by Majorana:
\begin{equation}
\label{eq:mass1}
\mu_\gamma =  
\frac{E+\unit{v}\bdot\grad\Phi}{\w_{p0}A(E+\unit{v}\bdot\grad\Phi) 
 + 4\pi\delta\dot{v}\big(n_0+\tilde{n}f(\ell,q)\big)}
\end{equation}
This expression can be written in a form that easily recalls the
original Majorana formula of \Eqn{eq:majmass},
\begin{equation}
\mu_\gamma =  \frac{m^\ast}{\Sigma(\ell,q)+ 1/2 },
\label{procamajorana}
\end{equation}
where $\Sigma(\ell,q)$ is a general function of the OAM of the photon
and of the characteristic spatial scale of the perturbation that
summarises all the terms present in \Eqn{eq:mass1}, and the Majorana mass
term coincides with that acquired by the photon in the unperturbed
plasma.

Unlike the mathematical structure of the space-time manifold described
by the Lorentz group, in which space is homogeneous and isotropic and
time homogeneous, a plasma may exhibit spatial/temporal structures
that break the space-time symmetry. The inhomogeneity of plasma
structures and the interaction between photons and charges in the
plasma actually change the Proca mass turning time invariance into a
more complicated space/time invariance, with the result of converting
part of the Proca mass into orbital angular momentum. Moreover, the
mathematical correspondence between the Majorana solution and the
behaviour of a photon in a plasma confirms the \emph{Ansatz} that
OAM states cannot induce a negative squared mass to the photon, if
Lorentz invariance is to be preserved. Alternatively, one can prove
this conjecture by following the relationship between Proca mass and
OAM in Ref.~\citenum{Tamburini&al:EPL:2010} in the simplest case of
an helicoidal perturbation: the mass squared term of the photon is
positive if the Proca mass is larger than the absolute value of the term
containing OAM,
\begin{multline}
 E \w^2_{p0}\big(1+\eta(r,\varphi,z)\big)
 + 4\pi\unit{v}\bdot\dalembert\grad\Phi
\\
  >4\pi\delta\dot{v}\big(n_0+\tilde{n}\cos(\ell\varphi+q_0z)\big)
\end{multline}
which is found to be always valid, in agreement with the hypothesis
of small perturbation assumed in our calculations. 

The Majorana Tower, derived from infinitesimal Lorentz transformations,
allows also solutions for which the energy $W$ and momentum $\p$ obey
a tachyonic relationship, for which $W= \pm\sqrt{c^2p^2 - k^2c^4}$.
This solution, which can be interpreted in terms of Lorentz invariance
violation \cite{Tamburini&Laveder:arXiv:2011}, is beyond the scope of
the present Letter.

\section{Conclusions}

Photons that propagate in a structured plasma acquire mass and orbital
angular momentum because of the hidden gauge invariance due to the
Anderson-Higgs mechanism in Proca-Maxwell equations. The squared-mass
term that describes photon OAM acts as a negative squared mass term
that could apparently cause photons to acquire a negative squared mass.
We find, however, that the mass/total angular momentum relationship of
Proca-Maxwell equations forbids the negative squared mass condition.

The presence of the plasma, and the spatial structuring with
characteristic scale length $q$ induce a mass/total angular momentum
relationship with the properties of the spin/mass relationship expected
for massive Majorana particles. The infinite spin Majorana's solution to
the Dirac equation gives, by definition, a spectrum of particles with
a positive definite or null finite mass. The difference between the
original solution proposed by Majorana, based on space-time symmetries
of the Lorentz group applied to the Dirac equation in free space, and
that of photons in a structured plasma lies in the contribution of the
OAM to the mass/angular momentum relationship in this latter case.

The presence of a characteristic scale length and structures in the
plasma, as expected in nature, shows clear analogies with models of
space-time involving modified action of the Lorentz group, such as
the Magueijo-Smolin model, in which there exists an energy scale
that acts as invariant \cite{Magueijo&Smolin:PRL:2002}. Moreover, the
mathematical structure found in our work shows deep analogies with
the dynamics described by the Dirac equation in the presence of a
characteristic scale that unavoidably introduces Berry phase effects
\cite{Gosselin&al:PLB:2008}.

This type of solution could find interesting applications in the
propagation of light and radio waves in plasma, in plasma physics and
more generally in studies of the propagation of photons in nonlinear
inhomogeneous media. Our result can be applied to other particles such
as neutrinos, involving spinor fields, and gravitons giving a new
perspective to the meaning of the Majorana Tower.

The extension of the Majorana Tower to gravitons is not so immediate.
In fact, until now, no theory of quantum gravity exists, and no
gravitational waves have been directly observed. The coupling of
gravitons and the plasma frequency is not known and could have a
different strength and even lead to imaginary effective masses, which
would correspond to the impossibility of screening gravitational waves
with plasma.

%

\end{document}